\documentclass[sigconf,natbib=true,anonymous=false]{acmart}

\usepackage{amssymb}
\usepackage{array}
\usepackage{url}
\usepackage{dblfloatfix}
\usepackage{booktabs}
\usepackage{amsmath}
\usepackage{tabularx}
\usepackage{colortbl}

\usepackage{graphics}
\usepackage{epsfig}
\usepackage{multirow}
\usepackage{flushend}

\usepackage{url}
\usepackage{tikz}
\usepackage{enumitem}
\usepackage{multicol}

\usepackage{verbatim}
\usepackage{float}
\usepackage{subcaption}

\usepackage{algorithm}
\usepackage{algpseudocode}

\usepackage{marginnote}
\usepackage{xspace}
\usepackage{balance}

\usepackage{todonotes}
\usepackage{xcolor}

\newcommand{\squishlist}{
 \begin{list}{$\bullet$}
  { \setlength{\itemsep}{0pt}
     \setlength{\parsep}{1pt}
     \setlength{\topsep}{1pt}
     \setlength{\partopsep}{0pt}
     \setlength{\leftmargin}{1.5em}
     \setlength{\labelwidth}{1em}
     \setlength{\labelsep}{0.5em} } }

\newcommand{\squishend}{
  \end{list}  }



\copyrightyear{2022} 
\acmYear{2022} 
\setcopyright{none}
\acmConference[AKBC '22]{
AKBC 2022 Workshop, Knowlege Graphs in Finance and Economics}{November 5, 2021}{London, UK}
\acmBooktitle{AKBC 2022 Workshop, Knowledge Graphs in Finance and Economics (AKBC Workshop '22), November 5, 2021, London, UK}

\author{Iain Mackie}
\affiliation{
  \institution{University of Glasgow}
}
\email{i.mackie.1@research.gla.ac.uk}

\author{Jeffrey Dalton}
\affiliation{
  \institution{University of Glasgow}
}
\email{jeff.dalton@glasgow.ac.uk}

\renewcommand\footnotetextcopyrightpermission[1]{}

\begin{document}
\fancyhead{}

\title{Query-Specific Knowledge Graphs for Complex Finance Topics}

\begin{abstract}

Across the financial domain, researchers answer complex questions by extensively ``searching'' for relevant information to generate long-form reports. 
This workshop paper discusses automating the construction of query-specific document and entity knowledge graphs (KGs) for complex research topics.
We focus on the CODEC dataset, where domain experts (1) create challenging questions, (2) construct long natural language narratives, and (3) iteratively search and assess the relevance of documents and entities.
For the construction of query-specific KGs, we show that state-of-the-art ranking systems have headroom for improvement, with specific failings due to a lack of context or explicit knowledge representation. 
We demonstrate that entity and document relevance are positively correlated, and that entity-based query feedback improves document ranking effectiveness.
Furthermore, we construct query-specific KGs using retrieval and evaluate using CODEC's ``ground-truth graphs'', showing the precision and recall trade-offs.
Lastly, we point to future work, including adaptive KG retrieval algorithms and GNN-based weighting methods, while highlighting key challenges such as high-quality data, information extraction recall, and the size and sparsity of complex topic graphs.

\end{abstract}

\maketitle

\section{Introduction}
\label{sec:intro}

\begin{figure*}[h!]
    \centering
    \includegraphics[scale=0.4]{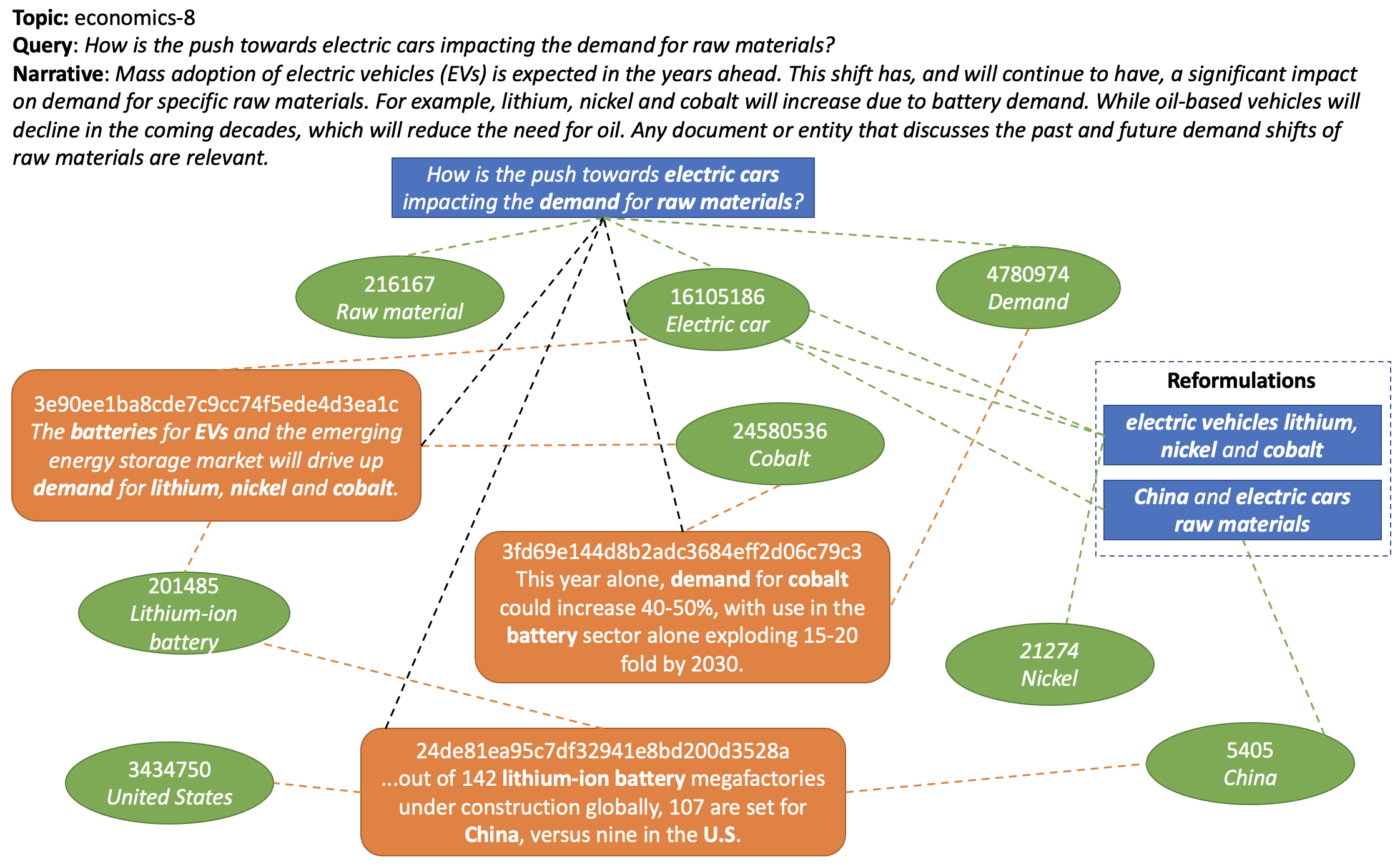}
    \caption{Free-text query-specific KG constructed for CODEC topic economics-8. Query nodes are in blue, document nodes are in orange, entity nodes are in green, and dotted lines represent entity links.}
    \label{fig:topic-graph}
\end{figure*}

Financial analysts research complex topics to identify relevant documents, key entities, and valuable facts.
For example, if a researcher wants to understand, \textit{How is the push towards electric cards impacting demand for raw materials?}. 
In practice, a researcher would use a search engine and reformulate queries to investigate dimensions of the topic, e.g. ``Electric vehicles lithium'', ``China and electric cards raw materials'', etc. 
Through this iterative process, the researcher identifies useful sources (i.e. documents) and understands the critical concepts (i.e. entities). 
Figure \ref{fig:topic-graph} shows an example query-specific KG that analysts would use to support their end goal, such as writing an investment report.


A standard approach to automatically construct query-specific free-text KGs is using retrieval systems to identify relevant documents and entities, and running entity linkers over the documents to extract structured connections \cite{dietz2019ent, dalton2013constructing, zhao2020complex}.
Additionally, neural, sparse and learning-to-rank (L2R) retrieval methods can provide query-specific relevance weightings to the KG nodes and edges  \cite{dietz2019ent, zhao2020complex}.
This paper focuses primarily on the retrieval component of query-specific KG construction, where in recent years, LLM-based re-ranking \cite{Li_etal_2020_PARADE,  MacAvaney_etal_SIGIR2019, nogueira2020document} and dense retrieval systems \cite{khattab2020colbert, xiong2020approximate, wu2019zero} dominate ranking leaderboards.
However, findings have shown failures on even simple entity queries \cite{sciavolino2021simple}, and headroom for improvement on complex queries where the answers are long and span multiple documents \cite{mackie2021dlhard}.

In this research, we use CODEC dataset \cite{mackie2022codec} that contains complex research questions and aligned document and entity relevance judgements.
We specifically focus on topics within the finance domain and show that current neural models fail due to missing context and a lack of an explicit knowledge representation. 
We find real-world finance questions are 10-15\% harder for entity and document ranking systems than for history or politics questions. 

Prior work has shown that ad-hoc retrieval \cite{Dalton2014EntityQF, xiong2017word} and entity ranking \cite{dietz2019ent} can be improved by leveraging entity information.
Using CODEC's aligned document and entity relevance judgments we can understand this relationship better for complex finance topics.
Specifically, that document relevance positively correlates with the proportion of critical entities contained within the document (+0.21 Pearson Correlation Coefficient) for finance topics. 
We leverage this relationship to improve initial retrieval with entity query expansion, improving Recall@1000 over BM25 by 22\% and 12\% of term-based RM3 expansion.  

We use strong document and entity retrieval systems to construct query-specific KGs for CODEC automatically. 
Specifically, we vary $k_{doc}$ document run depth to add more or fewer document nodes, and vary $k_{ent}$ entity run depth to filter entities and entity links based on a relevance threshold.
For evaluation, we use CODEC's entity and document judgments to develop ``ground-truth KGs''.
This evaluation shows the precision and recall trade-offs when varying different run depths, and headroom by incorporating graph-based on text-based methods. 
Furthermore, this analysis clearly highlights KGs for complex finance topics are much large compared to smaller QA-focused graphs \cite{zhao2020complex, yasunaga2021qa}, containing potentially hundreds of relevant nodes and thousands of relevant edges.

Lastly, we discuss future research directions and challenges. 
Specifically, we outline opportunities for improvements in graph construction leveraging both graph structure and LLM-based text embeddings. 
For example, new adaptive KG retrieval algorithms and GNN-based scoring models that build on prior work from QA \cite{zhao2020complex, yasunaga2021qa}. 
At the same time, we highlight vital challenges such as large datasets to train neural query-specific KG construction models and high-quality test data that contains ground truth edge annotations.
Additionally, we highlight challenges around information extraction effectiveness \cite{kadry2017open}, tail entities \cite{zhang2019long}, and dealing with the size and sparsity of complex topic graphs.

\section{CODEC}
\label{sec:codec}

Complex Document and Entity Collection 
(CODEC) \cite{mackie2022codec}\footnote{available at \url{https://github.com/grill-lab/CODEC} and \textit{ir-datasets} \cite{macavaney:sigir2021-irds}} is a resource that focuses on complex research topics. 
CODEC covers two tasks: document ranking and entity ranking
Therefore, for a given information need $Q$, systems need to return a ranked list of documents $[D_1, D_2..D_N]$ and a ranked list of entities $[E_1, E_2..E_N]$
CODEC contains 42 essay-style questions and corresponding long natural language narratives across finance, history and politics domains.
The document corpus is a 730k-focused web document 
collection of domain-specific content,  the entity KB is Wikipedia \cite{petroni2021kilt}, and the entity linking are used to ground documents.

In this work, we will focus solely on the finance topics subset that covers wide-ranking research areas, including public sector enterprises, the dot-com bubble, the EU's monetary union, and the NFT investment category. 
Domain experts (FX Trader, Investment Banker, Accountant, etc.) are employed to provide deep domain insights for high-quality annotations in this complex domain.
This subset includes 14 topics, 127 manual query formulations (over 9 per topic on average), long narrative answers that contain 114 words and 32 entities \cite{piccinno2014tagme} on average. 
Furthermore, these topics contain 3,568 entity (254.9 per topic) and 1,970 document relevance judgments (140.7 per topic).


\section{Systems Analysis}
\label{sec:analysis}

For the task of automatically constructing a query-specific free-text KG, we focus on the individual components of document and entity ranking.
These ranking components are critical as they will directly impact the quality of the KG, which can be explicitly evaluated via relevance judgments. 
We conduct systems analysis using strong sparse retrieval methods (BM25 \cite{robertson1994some} and BM25+RM3 \cite{abdul2004umass}), a dense retrieval model (ANCE \cite{xiong2020approximate}), and LLM-based re-ranker (T5 \cite{nogueira2020document}).

\subsection{Document ranking}

Table \ref{tab:document-baselines} shows system performance for document ranking. 
Based on Recall@1000 of 0.735 the best performing method is BM25+RM3, and T5 re-ranking improves MAP to 0.303 and NDCG@100 to 0.487. 
These measures indicate significant headroom for improvement when compared to other benchmarks \cite{MS_MARCO_v1}.
This headroom for retrieval effectiveness will be vital for the accurate construction of query-specific KGs.


\begin{table}[h!]
\caption{Document ranking effectiveness on CODEC finance topics.}
\label{tab:document-baselines}
\begin{tabular}{l|r|r|r|}
\cline{2-4}
                                              & \multicolumn{1}{l|}{\textbf{NDCG@100}} & \multicolumn{1}{l|}{\textbf{MAP}} & \multicolumn{1}{l|}{\textbf{Recall@1000}} \\ \hline
\multicolumn{1}{|l|}{\textbf{BM25}}           & 0.330                             & 0.161                                 & 0.702                                    \\ \hline
\multicolumn{1}{|l|}{\textbf{BM25+RM3}}       & 0.360               & 0.182                                & \textbf{0.735}             \\ \hline
\multicolumn{1}{|l|}{\textbf{ANCE-MaxP}}      & 0.349                             & 0.184                                & 0.656                                     \\ \hline
\multicolumn{1}{|l|}{\textbf{BM25+T5}}    & 0.486      & 0.301          & 0.702             \\ \hline
\multicolumn{1}{|l|}{\textbf{BM25+RM3+T5}}        &  \textbf{0.487}             & \textbf{0.303}                   & \textbf{0.735}                                     \\ \hline
\multicolumn{1}{|l|}{\textbf{ANCE-MaxP+T5}}   & 0.470               & 0.282                 & 0.656                                    \\ \hline
\end{tabular}
\end{table}

Furthermore, we use BM25+RM3+T5 to compare the hardness of finance topics to other domains present in CODEC (history and politics).
Table \ref{tab:domains} shows that finance is by far the most challenging domain for both initial retrieval and re-ranking.
An example of a hard finance topic is economics-12, \textit{What are the common problems or criticisms aimed at public sector enterprises?}, with Recall@1000 of under 0.55 for all systems.
This topic requires a lot of latent knowledge, and analysis of relevant documents shows they contain minimal keyword overlap with the query.
Additionally, neural re-ranking systems fail on economic-21 topic, \textit{How much of a threat are ETFs to actively-managed Asset Managers?}, with BM25+RM3+T5 having a MAP of only 0.138.
The system runs show that many highly ranked documents do contain entities [ETFs] or [Asset Managers], but current models cannot filter based on the relationships that explain the ``threat'' ETFs pose to Asset Managers.
This motivates more advanced structure reasoning within neural ranking methods.

\begin{table}[h!]
\caption{CODEC document ranking effectiveness by domain (system: BM25+RM3+T5).}
\label{tab:domains}
\begin{tabular}{l|r|r|r|}
\cline{2-4}
                                            & \multicolumn{1}{l|}{\textbf{NDCG@100}} & \multicolumn{1}{l|}{\textbf{MAP}} & \multicolumn{1}{l|}{\textbf{Recall@1000}} \\ \hline
\multicolumn{1}{|l|}{\textbf{History}}     & 0.597                             & 0.391                                & 0.850                                      \\ \hline
\multicolumn{1}{|l|}{\textbf{Politics}}     & 0.542                             & 0.344                                & 0.816                                      \\ \hline
\multicolumn{1}{|l|}{\textbf{Finance}}    & 0.487                              & 0.303                                 &  0.735             \\ \hline 
\end{tabular}
\end{table}

Lastly, system failures show that low document ranking effectiveness was strongly related to missing documents that contain key topic concepts. 
For example, all systems had a MAP under 0.150 for topic economics-2, \textit{What technological challenges does Bitcoin face to becoming a widely used currency?}.
Highly ranked documents contain anticipated named entities, i.e. [Bitcoin], [Blockchain], and [Satoshi Nakamoto]. 
However, systems miss the key concepts that are needed to truly understand this information need, i.e. [Transaction time], [Transaction cost], [Quantum technology], [Carbon footprint], and [Cyberattack].

\subsection{Entity ranking}

Table \ref{tab:entity-baselines} shows the system performance on the entity ranking task. 
The performance of entity ranking systems is much lower when compared to document ranking, emphasising that entity ranking is a challenging task within the CODEC setup, i.e. searching for concepts as well as named entities.
Neural ranking methods (ANCE and T5) significantly underperform sparse methods (BM25 and BM25+RM3), at least in a zero-shot setting, and we did not include results. 
The best system is BM25+RM3 with Recall@1000 of 0.651, which is 13\% improvement over BM25, showing the clear benefits of pseudo-relevance feedback within entity ranking of complex finance topics.




\begin{table}[h!]
\caption{Entity ranking effectiveness on CODEC finance topics.}
\label{tab:entity-baselines}
\begin{tabular}{l|r|r|r|}
\cline{2-4}
                                              & \multicolumn{1}{l|}{\textbf{NDCG@100}} & \multicolumn{1}{l|}{\textbf{MAP}} & \multicolumn{1}{l|}{\textbf{Recall@1000}} \\ \hline
\multicolumn{1}{|l|}{\textbf{BM25}}           & 0.325                             & 0.146                                 & 0.575                                    \\ \hline
\multicolumn{1}{|l|}{\textbf{BM25+RM3}}       & \textbf{0.352}               & \textbf{0.177}                                & \textbf{0.651}             \\ \hline

\end{tabular}
\end{table}

Similar to the findings for document ranking, Table \ref{tab:domains-entity} supports that finance topics are more challenging for entity ranking systems than history and politics topics.
This highlights the complexities of the finance domain, which require deep contextualisation and reasoning to answer real-world finance questions. 
Analysing specific topic failures, for example, economics-17 topic ``Why is scaling a hardware business more capital intensive than a software business?''. 
We see that retrieval models cannot identify relevant entities that are categories or sub-categories of [Hardware] or [Software] businesses related to [Capital expenditure], which requires more extensive multi-hop reasoning capabilities.




\begin{table}[h!]
\caption{CODEC entity ranking effectiveness by domain (system: BM25+RM3).}
\label{tab:domains-entity}
\begin{tabular}{l|r|r|r|}
\cline{2-4}
                                            & \multicolumn{1}{l|}{\textbf{NDCG@100}} & \multicolumn{1}{l|}{\textbf{MAP}} & \multicolumn{1}{l|}{\textbf{Recall@1000}} \\ \hline
\multicolumn{1}{|l|}{\textbf{History}}     & 0.409                             & 0.224                                & 0.656                                      \\ \hline
\multicolumn{1}{|l|}{\textbf{Politics}}     & 0.415                             & 0.226                                & 0.747                                      \\ \hline
\multicolumn{1}{|l|}{\textbf{Finance}}    & 0.352                              & 0.177                                 &  0.651             \\ \hline 
\end{tabular}
\end{table}

\section{Document and Entity Relationship}
\label{sec:doc-ent}

For entity linking, we run REL \cite{vanHulst:2020:REL}  entity linker over the entire CODEC 730k document corpus to provide structured connections between documents and entities, enabling construction of a query-specific graph.
This results in 27.5m entity links (37.7 entity links per document).
To understand the relationship between document relevance and entity relevance for the finance domain, we connect the document judgments and the relevance of entities mentioned in each document using the entity links and entity judgments.
We calculate the Pearson Correlation Coefficient between document relevance and the percentage of entities in the document grouped by entity relevance.
We find that entity links (+0.21) support that documents with higher proportions of highly relevant entities are positively correlated with document relevance.



\begin{table}[h!]
\caption{Entity-QE document ranking effectiveness against initial retrieval systems.}
\label{tab:entity-qe}
\begin{tabular}{l|r|r|r|}
\cline{2-4}
                                            & \multicolumn{1}{l|}{\textbf{NDCG@100}} & \multicolumn{1}{l|}{\textbf{MAP}} & \multicolumn{1}{l|}{\textbf{Recall@1000}} \\ \hline
\multicolumn{1}{|l|}{\textbf{BM25}}     & 0.330                             & 0.161                                & 0.702                                      \\ \hline
\multicolumn{1}{|l|}{\textbf{BM25+RM3}}     & 0.360                             & 0.182                                & 0.735                                      \\ \hline
\multicolumn{1}{|l|}{\textbf{Entity-QE}}    & \textbf{0.404}                              & \textbf{0.220}                                 & \textbf{0.799}             \\ \hline
\end{tabular}
\end{table}

We build on these findings by using an oracle entity feedback method that enriches the query with names of relevant entities taken from entity judgments, Entity-QE.
Tables \ref{tab:entity-qe} shows that Entity-QE improves Recall@1000 over BM25 by 22\% and 12\% of term-based RM3 expansion. 
This is an example of how co-optimising both document and entity ranking offers many opportunities to improve the construction of query-specific KGs. 


\section{Query-Specific KGs}

We define a simple graph construction approach based on document and entity ranking and entity linking, 
similar to prior successful query-specific KG approaches \cite{dietz2019ent, zhao2020complex}.
Specifically, given the user's query $Q$, we retrieve top-ranked $k_{doc}$ set of documents ${[D_1,D_2,D_3...,D_{k_{doc}}]}$ and $k_{ent}$ set of entities ${[E_1,E_2,E_3...,E_{k_{ent}}]}$.
The top-ranked documents are fed into the entity linking pipeline, which provides entity links ${E_1,E_2,...,E_n} \in D$. 
Lastly, to reduce the number of non-topic-relevant entities in our KG, we use entity ranking to filter entities. 

Therefore, we create a query-specific graph $G = (V, E)$, where $V$ are a set of vertices (nodes) of entity, document, or query types.
Furthermore, $E$ is a set of edges based on entity links (i.e. document entity mentions) and retrieval query-documents and query-entity edges.
The graph is inherently query-specific as we only contain the more relevant document and entities based on system runs.

\subsection{Ground-Truth KGs}

To theoretically ground the evaluation of graph quality for query-specific KGs, we define the ``ground-truth KGs" for each topic.
Doing so will allow us to measure the differences between this ground truth graph and automatically constructed graphs.
This evaluation is possible using CODEC as we have ground-truth query-specific entity and document judgments.
However, we assume WAT document-level entity links are ``correct'', which is not always the case, and follow-up work should address this.
Specifically, we define the topic ground-truth KGs as the query connected to all relevant document and entity nodes, and includes document-entity edges based on entity mentions where the entity is itself relevant.  
We use the CODEC binary relevance judgments.

\begin{table}[h!]
\caption{Node and edge statistics of CODEC ground-truth KGs.}
\label{tab:golden-stats}
\begin{tabular}{l|l|}
\cline{2-2}
                                         & \textbf{Mean Count} \\ \hline
\multicolumn{1}{|l|}{\textbf{Query Nodes}}   & 1.00                     \\  \hline
\multicolumn{1}{|l|}{\textbf{Document Nodes}}   & 44.1                    \\  \hline
\multicolumn{1}{|l|}{\textbf{Entity Nodes}}   & 45.5                      \\  \hline
\multicolumn{1}{|l|}{\textbf{Query-Document Edges}} & 44.1               \\  \hline
\multicolumn{1}{|l|}{\textbf{Query-Entity Edges}} & 45.5              \\  \hline
\multicolumn{1}{|l|}{\textbf{Document-Entity Edges}} & 1,227.1              \\ \hline
\end{tabular}
\end{table}

Table \ref{tab:golden-stats} shows the ground-truth graph node and edge statistics across the 14 finance CODEC topics.
These graphs are relatively precision-focused based on judged relevant documents and linked relevant entities, with 44.1 document nodes and 45.5 entity nodes on average. 
The ground-truth graphs, on average, have 1,227.1 document-to-entity edges, showing that the relevant entities are central to these topics (approximately 50 edges per entity).
However, based on WAT entity links, only 25.2 entity nodes are directly connected to relevant documents (only 55\% of total relevant entities), highlighting entity linking failings or many useful named entities and concepts in relevant documents.

When comparing automatically constructed KGs to our ground-truth graphs, we focus on evolution two-fold: (1) precision and (2) recall.  
Specifically, we define precision as the proportion of nodes or edges present in constructed KGs when comparing them to the ground-truth graphs, $P_{node} = |KG_{truth\_nodes}| / |KG_{nodes}|$, this is propositional to works across different graph sizes.
This metric focuses on graph scarcity, a significant issue in the accuracy of finance tasks using the KG, for example, factually-grounded text generation or QA. 
Additionally, we define recall as the metric for the proportion of nodes or edges missing from the constructed KGs when comparing them to the ground-truth graphs, $R_{node} = | KG_{truth\_nodes}| / |Ground Truth_{nodes}|$).
In finance, recall measures the quality of missing relevant information, which is vital when use cases require ``knowing all the facts'', for example, investment due diligence.

\subsection{Experiments}

We evaluate automatic KGs using CODEC's ground-truth graphs and strong system runs.
For example, using the most effective document ranking system, BM25+RM3+T5, we can investigate different KG properties as we vary  $k_{doc}$.
We can also use the most effective entity ranking system, BM25+RM3, and varying $k_{ent}$ to vary the automatic exclusion of query and document entity mentions based on relevance.
For entity linking, we use a WAT \cite{piccinno2014tagme} wikification system to provide high-recall named entities and concept connections.
This results in 116.5m entity links across the document corpus (159.6 entity links per document), which equated to roughly one entity link every 4 terms.

Table \ref{tab:precision} and \ref{tab:recall} show precision and Recall trade-offs of automatic KG construction as we vary $k_{doc}$ and  $k_{ent}$ across [10,50,100,250,1000].
We specifically report graph quality results for document nodes, entity nodes, query-document edges, query-entity edges, and document-entity edges.

\begin{table}[h!]
\caption{Precision of automatic graph with $k_{doc}$, $k_{ent}$ = [10, 50, 100, 250, 1000]}
\label{tab:precision}
\begin{tabular}{l|l|l|l|l|l|}
\cline{2-6}
                                                     & \textbf{10} & \textbf{50} & \textbf{100} & \textbf{250} & \textbf{1000} \\ \hline
\multicolumn{1}{|l|}{\textbf{Document Nodes}}         & 0.56  & 0.33           & 0.22            & 0.11            & 0.03             \\ \hline
\multicolumn{1}{|l|}{\textbf{Entity Nodes}}         & 0.38 & 0.23           & 0.16           & 0.09           & 0.03             \\ \hline
\multicolumn{1}{|l|}{\textbf{Query-Document Edges}}        & 0.56  & 0.33           & 0.22            & 0.11            & 0.03             \\ \hline
\multicolumn{1}{|l|}{\textbf{Query-Entity Edges}}         & 0.38 & 0.23           & 0.16           & 0.09           & 0.03             \\ \hline
\multicolumn{1}{|l|}{\textbf{Document-Entity Edges}} & 0.38 & 0.28           & 0.18            & 0.09            & 0.02             \\ \hline
\end{tabular}
\end{table}


As expected, when $k_{ent}$ and $k_{doc}$ are at lower thresholds, precision is significantly higher compared to the larger thresholds.
For example, when $k_{ent}$ and $k_{doc}$ are equal to 10, the precision of document nodes is 0.56 and entity nodes is 0.62.
However, the document-entity edges only have a precision of 0.38, the compounding of errors due to included non-relevant documents and entities.
Nonetheless, compared to recall of  $k_{ent}$ and $k_{doc}$ equal 10, around 90\% of nodes and 95 \% or edges are missing, highlighting the need for large graphs to capture the domain complexities of complex finance topics. 

\begin{table}[h!]
\caption{Recall of automatic graph with $k_{doc}$, $k_{ent}$ = [10, 50, 100, 250, 1000]}
\label{tab:recall}
\begin{tabular}{l|l|l|l|l|l|}
\cline{2-6}
                                                     & \textbf{10} & \textbf{50} & \textbf{100} & \textbf{250} & \textbf{1000} \\ \hline
\multicolumn{1}{|l|}{\textbf{Document Nodes}}        & 0.13 & 0.37           & 0.50            & 0.63            & 0.73             \\ \hline
\multicolumn{1}{|l|}{\textbf{Entity Nodes}}          &  0.08 & 0.25           & 0.36            & 0.49            & 0.65             \\ \hline
\multicolumn{1}{|l|}{\textbf{Query-Document Edges}}        & 0.13 & 0.37           & 0.50            & 0.63            & 0.73             \\ \hline
\multicolumn{1}{|l|}{\textbf{Query-Entity Edges}}          &  0.08 & 0.25           & 0.36            & 0.49            & 0.65             \\ \hline
\multicolumn{1}{|l|}{\textbf{Document-Entity Edges}} & 0.05 & 0.23           & 0.33            & 0.50            & 0.67             \\ \hline
\end{tabular}
\end{table}

Around $k_{ent}$ and $k_{doc}$ equal to 50, the precision and recall trends cross each other, with document nodes scoring 0.35, entity nodes scoring 0.25, and document-entity edges scoring 0.25.
This highlights the relatively low-performing graph construction, driven by two-fold: (1) this is a challenging task incorporating multiple systems and (2) more advanced methods are required to improve performance. 
Specifically, future work will focus on incorporating more graph-based pruning techniques, the performance of entity linking techniques \cite{kadry2017open}, and graphs-information being used as part of the retrieval and re-ranking processes.  

When $k_{ent}$ and $k_{doc}$ are increased to 1000, recall for documents is 0.73, entities are 0.65, and edges are 0.67.
Although recall improves as $k$ increases, there are many missing graph nodes and edges.
Specifically, 27\% of document nodes and 35\% of entity nodes are missing, which highlights the opportunities to improve the neural re-ranking of documents and entities on complex finance topics.

Overall, this analysis shows that the CODEC dataset can be effectively used to benchmark automatic knowledge graph construction on complex finance topics. 
We show that simple methods can construct KGs and how ground-truth KGs can evaluate construction performance, highlighting specific areas for improvement.
Lastly, we show that the automatic graphs constructed for complex finance topics can be very large, with high-recall KGs having 1k+ of nodes and 100k+ edges.

\section{Future Work}
\label{sec:fw}

We plan to extensively research automated graph construction on complex topics, including a focus on the finance domain due to the domain complexity. 
Specifically, building upon recent adaptive retrieval work \cite{macavaney2022adaptive}, we plan to leverage entities and graph-based information to improve system recall.
Additionally, building on successful applications with factoid and multiple choice QA \cite{zhao2020complex, yasunaga2021qa}, we aim to develop GNN-based scoring models for complex research topics, which will leverage both LLM embeddings and graph-based information.
However, these examples in QA often only contain a single correct node and tens or low hundreds of additional context nodes.
Conversely, complex finance topics can contain several queries, a hundred relevant document and entity nodes, and thousands of context nodes. 
This scale and potential sparsity will be a specific challenge for GNN-based model training.  

Other long-term challenges will also require addressing.
For example, the performance of information extraction systems has a significant impact on graph quality \cite{kadry2017open}, with specific failing on tail entities \cite{zhang2019long}.
For example, we find that 40\% of the relevant entities were not linked to any relevant document. 
Additionally, high-quality data capturing the ``ground-truth facts'' is required for more in-depth system evaluation.
Currently, we assume a relevant edge between all relevant documents and entities mentioned within that document.
Therefore, we will run a user study to generate detailed fact-based data.


\section{Acknowledgements}
\label{sec:ack}

We would like to acknowledge all the authors on the original CODEC paper (Paul Owoicho, Carlos Gemmell, Sophie Fischer, and Sean MacAvaney) and Ivan Sekulić for on-going collaboration on query-specific KGs.
This work is supported by the 2019 Bloomberg Data Science Research Grant and the Engineering and Physical Sciences Research Council grant EP/V025708/1. 





\bibliographystyle{ACM-Reference-Format}
\balance
\bibliography{foo}

\end{document}